\documentclass[pra,superscriptaddress,twocolumn,amsmath,amssymb]{revtex4}
\usepackage{graphicx}
\usepackage{bm}
\usepackage{amsmath,amssymb}
\usepackage{color,multirow}
\include{Qcircuit}

\newtheorem{lemma}{Lemma}
\newtheorem{theorem}{Theorem}

\newcommand{\sX}{\mathsf{X}}
\newcommand{\sZ}{\mathsf{Z}}
\newcommand{\sW}{\mathsf{W}}
\newcommand{\bx}{{\bf x}}
\newcommand{\bz}{{\bf z}}
\newcommand{\bC}{\mathbb{C}}
\newcommand{\bF}{\mathbb{F}}

\newcommand{\Tr}{{\rm Tr}\,}

\def\QED{\mbox{\rule[0pt]{1.5ex}{1.5ex}}}

\def\endproof{\hspace*{\fill}~\QED\par\endtrivlist\unskip}

 \newenvironment{proofof}[1]{\vspace*{5mm} \par \noindent
         \quad{\it Proof of #1:\hspace{2mm}}}{\endproof
}

\begin{document}
\title{Verifiable measurement-only blind quantum computing with stabilizer testing}
\author{Masahito Hayashi}
\affiliation{Graduate School of Mathematics, Nagoya University,
Furocho, Chikusa-ku, Nagoya, 464-860, Japan}
\affiliation{Centre for Quantum Technologies, National University of Singapore,
117543, Singapore}
\author{Tomoyuki Morimae}
\affiliation{ASRLD Unit, Gunma University, 1-5-1 Tenjincho, Kiryu-shi, Gunma, 376-0052, Japan}

\begin{abstract}
We introduce a simple protocol for verifiable measurement-only 
blind quantum computing. Alice, a client, can perform only single-qubit measurements,
whereas Bob, a server, can generate and store entangled many-qubit states.
Bob generates copies of a graph state, which is a universal resource state for measurement-based 
quantum computing, and sends Alice each qubit of them one by one.
Alice adaptively measures each qubit according to her program.
If Bob is honest, he generates the correct graph state, and therefore
Alice can obtain the correct computation result.
Regarding the security, whatever Bob does, Bob cannot learn any information about Alice's computation 
because of the no-signaling principle. 
Furthermore, malicious Bob does not necessarily send the copies of the correct graph state,
but Alice can check the correctness of Bob's state by 
directly verifying stabilizers of some copies.
\end{abstract}

\date{\today}
\maketitle  

\begin{figure}[htbp]
\begin{center}
\includegraphics[width=0.35\textwidth]{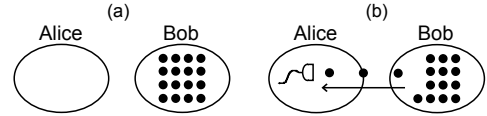}
\end{center}
\caption{
The measurement-only blind quantum computing.
(a) Bob generates a resource state.
(b) Bob sends Alice
each qubit of the resource state one by one. Alice adaptively measures 
each qubit.
} 
\label{fig0}
\end{figure}

Blind quantum computing is a quantum cryptographic protocol
that enables Alice (a client), who does not have any 
sophisticated quantum technology,
to delegate her quantum computing to Bob (a server), who
has a sufficiently powerful quantum computer, without leaking any her privacy.
The first protocol of blind quantum computing
that uses the measurement-based quantum computing~\cite{MBQC}
was proposed by Broadbent, Fitzsimons, and Kashefi~\cite{BFK},
and a proof-of-principle experiment was demonstrated with photonic qubits~\cite{Barz}.
In the protocol of Ref.~\cite{BFK}, 
Alice generates many randomly-rotated single-qubit states, 
and sends them to Bob.
Bob generates a universal resource state of the measurement-based quantum computing
by applying entangling gates on qubits sent from Alice.
Then, they do two-way classical communications:
Alice instructs Bob how to measure each qubit, and Bob returns
measurement results so that Alice can perform the feed-forward calculations.
It was shown in Ref.~\cite{BFK} that if Bob is honest, Alice can obtain the correct quantum
computing result (which we call the correctness), and that whatever evil Bob 
does, he cannot learn anything about Alice's input, output, and program 
(which we call the blindness)~\cite{unavoidable}. 
(See also Ref.~\cite{Vedran_composability} for a precise proof of the security.)
Inspired by the seminal result, plenty of improvements have been 
done~\cite{MABQC,Vedran_coherent,FK,AKLTblind,topoblind,topoveri,CVblind,Lorenzo,Joe_intern,tri,Sueki,distillation,
DI_Joe,DI_Elham,Carlos}.
For example, it was shown that instead of single-qubit states generation,
single-qubit measurements~\cite{MABQC} or coherent states generation~\cite{Vedran_coherent} 
are sufficient for Alice.
In the protocol of Ref.~\cite{MABQC}, so called the measurement-only
blind quantum computing, Bob generates a universal resource state of
measurement-based quantum computing (Fig~\ref{fig0}(a)), 
and sends each qubit of the resource state one by one to Alice (Fig.~\ref{fig0}(b)). 
Alice adaptively measures each qubit according to her program (Fig.~\ref{fig0}(b)).
Since adaptive single-qubit measurements on certain states 
are universal~\cite{MBQC,RHG,Gross,Miyake},
Alice with only single-qubit measurements ability can perform universal quantum computing
if Bob prepares the correct resource state.
Furthermore, since this protocol is a one-way quantum communication from Bob to Alice,
the blindness is guaranteed by the no-signaling principle~\cite{MABQC}.
Here, the no-signaling principle is one of the most fundamental
assumptions in physics, which says that if Alice and Bob share a system
she cannot transmit any her message to Bob whatever they do on their
systems. Quantum physics respects the no-signaling principle.

In addition to the correctness and the blindness,
the verifiability is another important requirement for blind quantum computing.
The verifiability means that Alice can check the correctness of Bob's computation.
Although the blindness guarantees that Alice's privacy is kept secret against 
malicious Bob,
it does not guarantee the correctness of the computation result with malicious 
Bob:
Bob cannot learn Alice's secret, but he can mess up the computation.
In order to avoid being palmed off a wrong result,
Alice needs some statistical test to verify the correctness
of Bob's computing.
There are several protocols that enable verifiable blind quantum 
computing~\cite{FK,topoveri,Matt,Vazirani,DI_Joe,DI_Elham}.
Some of them~\cite{Matt,Vazirani,DI_Elham} elegantly achieve the completely classical client, but
a trade-off is the requirement of more than two servers who do not communicate with
each other.
Although pursuing the completely classical client is an important direction,
in particular, for the goal of constructing an
interactive proof of BQP, where the assumption of non-communicating multi provers is natural,
in this paper we restrict ourselves to the single-server setup
assuming some minimum quantum technologies for the client,
since in the context of blind quantum computing,
assuming some minimum quantum technologies for the client is
more realistic than to assume that the client can verify that remote servers are not communicating
with each other. These results also achieve the
device independence. Although our protocol assumes the correctness
of measurement devices, it enables to derive a more practical bound
suitable for experiments.
Protocols in Refs.~\cite{FK,topoveri,DI_Joe} need only a single server
by assuming some minimum quantum technologies, which are available in
today's laboratories, for the client.
(The protocol of Ref.~\cite{FK} requires single-qubit states generations,
and those of Refs.~\cite{topoveri,DI_Joe} require single-qubit measurements
for the client.)
The idea of the verification in the protocols of Refs.~\cite{FK,topoveri,DI_Joe,DI_Elham} 
is to use trap qubits:
Alice secretly hides trap qubits in the resource state, and 
any disturbance of a trap signals Bob's dishonesty~\cite{FK,topoveri,DI_Joe,DI_Elham}.
An experimental demonstration of the idea was done with photonic qubits~\cite{BarzNP}.

\begin{figure}[htbp]
\begin{center}
\includegraphics[width=0.3\textwidth]{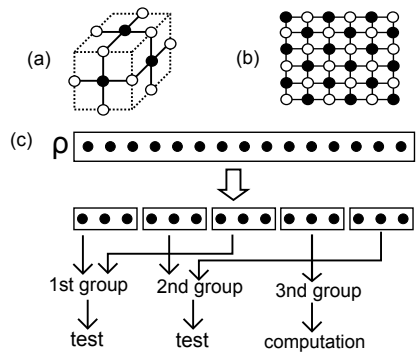}
\end{center}
\caption{
(a) The RHG lattice. (b) An example of bipartite graphs: the two-dimensional square lattice. 
Black and white colors indicate the bipartitions $B$ and $W$, respectively.
(c) An example for $n=3$, $k=2$. Two blocks go to the first group
and the other two blocks go to the second group. The left block goes to the third group.
} 
\label{fig}
\end{figure}

In this paper, we propose another protocol for verifiable 
measurement-only blind quantum computing.
The blindness is again guaranteed by the no-signaling principle
like Ref.~\cite{MABQC}.
The verifiability is, on the other hand,
achieved in a more straightforward way:
instead of hiding traps,
Alice directly checks whether the state sent from Bob is correct or not
by testing stabilizers~\cite{respect_Matt}.
Alice asks Bob to generate
$2k+1$ copies $|G\rangle^{\otimes 2k+1}$ of the graph state $|G\rangle$,
where $|G\rangle$ is an $n$-qubit graph state and $k=poly(n)$.
The graph state $|G\rangle$ is defined by
$
|G\rangle\equiv\Big(\bigotimes_{e\in E}CZ_e\Big)|+\rangle^{\otimes n},
$
where $|+\rangle\equiv\frac{1}{\sqrt{2}}(|0\rangle+|1\rangle)$, 
$E$ is the set of edges of $G$, and $CZ_e$ is the Controlled-$Z$ gate,
$CZ\equiv|0\rangle\langle0|\otimes I+|1\rangle\langle1|\otimes Z$, acting on
the pair of vertices sharing the edge $e$. 
The graph state $|G\rangle$ has the 
stabilizers
$
X_j\bigotimes_{i\in N(j)}Z_i,
$
for $j=1,2,...,n$,
where $N(j)$ is the set of the vertices connected to $j$.
Alice uses randomly chosen $2k$ copies of $|G\rangle^{\otimes 2k+1}$ to check
stabilizers, and the rest of it for her computation.
If Bob is honest, he generates $|G\rangle^{\otimes 2k+1}$, and in this case
we will show that she passes the test with probability 1.
If Bob is evil, on the other hand, he might generate another $n(2k+1)$-qubit state.
However, we will show
that if she passes the test,
the closeness of the single copy to the correct graph state 
$|G\rangle$ is guaranteed with a sufficiently small significance level. 
Any graph state can be used for our protocol as long as the corresponding
graph $G$ is bipartite. Therefore, for example, Alice can perform
the fault-tolerant topological measurement-based quantum computing~\cite{RHG} by taking
$|G\rangle$ as the 
Raussendorf-Harrington-Goyal
lattice~\cite{RHG} (Fig.~\ref{fig}(a)).

Note that there are several proposals for testing quantum gate 
operations~\cite{circuit1,circuit2},
but testing quantum circuit models 
assumes the identical and independent properties of each gate,
and suffers from the scalability and complexity of the analysis.
On the other hand, our result in the present paper 
(and Ref.~\cite{Matt}) demonstrate that
testing quantum computing becomes much easier if we consider a
measurement-based
quantum computing model,
which is a new interesting advantage of the measurement-based 
quantum computing model over the circuit model.
For more details about the relations between our result and previous works,
see Appendix.

{\it Protocol}.---
Our protocol runs as follows:
\begin{itemize}
\item[1.]
Honest Bob generates $|G\rangle^{\otimes 2k+1}$, where $|G\rangle$ is
an $n$-qubit graph state on a bipartite graph $G$,
whose vertices are divided into two disjoint sets $W$ 
and $B$. (Fig.~\ref{fig}(a) and (b).)
Bob sends each qubit of it one by one to Alice.
Evil Bob can generate any $n(2k+1)$-qubit state $\rho$
instead of $|G\rangle^{\otimes 2k+1}$.

\item[2.]
Alice divides $2k+1$ blocks of $n$ qubits into three groups by random choice.
(Fig.~\ref{fig}(c).)
The first group consists of $k$ blocks of $n$ qubits.
The second group consists of $k$ blocks of $n$ qubits.
The third group consists of a single block of $n$ qubits.

\item[3.]
Alice uses the third group for her computation.
Other blocks are used for the test, which will be explained later.
(Fig.~\ref{fig}(c).)
\item[4.]
If Alice passes the test, she accepts the result of the computation performed
on the third group.
\end{itemize}

\begin{figure}[htbp]
\begin{center}
\includegraphics[width=0.25\textwidth]{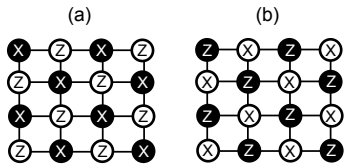}
\end{center}
\caption{
An example for the two-dimensional square lattice.
The measurement pattern for the first group (a)
and the second group (b).
} 
\label{fig2}
\end{figure}

For each block of the first and second groups,
Alice performs the following test:
\begin{itemize}
\item[1.]
For each block of the first group,
Alice measures qubits of $W$ in the $Z$ basis
and qubits of $B$ in the $X$ basis.
(Fig.~\ref{fig2}(a).)
\item[2.]
For each block of the second group,
Alice measures 
qubits of $B$ in the $Z$ basis
and
qubits of $W$ in the $X$ basis.
(Fig.~\ref{fig2}(b).)
\item[3.]
If the measurement outcomes in the $X$ basis
coincide with the values predicted from the outcomes in the $Z$ basis
(in terms of the stabilizer relations),
then the test is passed.
If any outcome in the $X$ basis
that violates the stabilizer relations is obtained,
Alice rejects.
\end{itemize}

{\it Analysis}.---
Let us analyze the correctness, blindness, and verifiability of our protocol.
First, our protocol is a one-way quantum communication from Bob to Alice,
and therefore, the blindness is guaranteed by the no-signaling principle
as in the protocol of Ref.~\cite{MABQC}.
Second, it is obvious that if $\rho=|G\rangle\langle G|^{\otimes 2k+1}$,
then Alice passes the test with probability 1. Therefore, if Bob is honest,
Alice passes the test with probability 1 and she obtains the correct computation
result on the third group. Hence the correctness is satisfied.
Finally, to study the verifiability, we consider
the following theorem:
\begin{theorem}
\label{L1}
Assume that $\alpha > \frac{1}{2k+1}$.
If the test is passed, 
with significance level $\alpha$,
we can guarantee that 
the resultant state $\sigma$ of the third group satisfies
\begin{eqnarray}
\langle G|\sigma|G\rangle
\ge 
1 -\frac{1}{\alpha(2k+1)}.\label{X1}
\end{eqnarray}
\end{theorem}
(Note that the significance level is
the maximum passing probability 
when malicious Bob sends incorrect states 
so that the resultant state $\sigma$ does not satisfy \eqref{X1}~\cite{textbook}.)
The proof of the theorem is given below and in Appendix.
From the theorem and the relation between the fidelity and trace norm \cite[(6.106)]{HIKKO}, 
we can conclude the verifiability:
If Alice passes the test, 
she can guarantee 
\begin{eqnarray*}
\Big|\mbox{Tr}(C\sigma)-\mbox{Tr}(C|G\rangle\langle G|)\Big|\le
\frac{1}{\sqrt{\alpha(2k+1)}}
\end{eqnarray*}
for any POVM $C$
with the significance level $\alpha$.
If we take $\alpha=\frac{1}{\sqrt{2k+1}}$, for example,
the left-hand side of the above inequality is
$\frac{1}{(2k+1)^{1/4}}\to0$ if $k\to \infty$,
and therefore the verifiability is satisfied.
Note that the lower bound, $\alpha>\frac{1}{2k+1}$, of the significance level $\alpha$
is tight, since
if Bob generates $2k$ copies of the correct state $|G\rangle $ and 
a single copy of a wrong state, 
Bob can fool Alice with probability $\frac{1}{2k+1}$,
which corresponds to $\alpha=\frac{1}{2k+1}$.

{\it Proof of Theorem}.---
The proof of the theorem is based on several interesting insights:
\begin{itemize}
\item[1.]
By considering an appropriate subspace,
we can reduce the problem to the test of a maximally-entangled state.
\item[2.]
For the test of a maximally-entangled state, 
verifications of coincidences of $X$ measurement results 
with $Z$ measurement results
are sufficient. Furthermore,
since we are interested in the fidelity between the given state and a maximally-entangled state,
we can consider, without loss of generality, the discretely twirled version of 
the given state, which drastically simplifies the problem~\cite{BDSW}.
\item[3.]
Finally, since we check the coincidence or discrepancy of the measurement results between two parties of the given bipartite cut, 
we have only to consider a distribution on $(0,0)$, $(0,1)$, $(1,0)$, and
$(1,1)$ for each block,
and therefore we can reduce the problem to a classical hypothesis testing.
\end{itemize}

Let us explain the first point.
Employing suitable classical data conversions, we can assume the following.
The systems ${\cal H}_B$ and ${\cal H}_W$ are written as
${\cal K}_B\otimes {\cal K}_B'$ and ${\cal K}_W\otimes {\cal K}_W'$
by using 
an $n_B'$-qubit system ${\cal K}_B$ and an $n_W'$-qubit system ${\cal K}_W$, 
respectively.
We denote the eigenstate corresponding to the eigenvalue all $0$ of $X$'s in ${\cal K}_B'$
by $|+\rangle_{B'}$, which is the graph state with isolated sites with no edge.
Similarly, we define $|+\rangle_{W'}$.
So, we find that the systems ${\cal K}_B$ and ${\cal K}_W$ are the same 
dimension, i.e., $n_B'=n_W'$.
Let $|G'\rangle$ be the graph state on ${\cal K}_B \otimes {\cal K}_W$ whose graph is composed of isolated edges.
The true state is given as the state $|G'\rangle \otimes |+\rangle_{B'} \otimes |+\rangle_{W'}$.
In this way, we can reduce the problem to that of the maximally-entangled state.
Note that Alice's measurements 
on ${\cal H}_B$ and ${\cal H}_W$
are replaced by 
on ${\cal K}_B$ and ${\cal K}_W$, respectively.
Applying the original Alice's measurement, Alice can realize the above modified 
measurement.
The detail of this discussion is given in Appendix.

Now let us explain the second point.
We focus on the 
Hilbert space $({\cal K}_B \otimes {\cal K}_W)^{\otimes (2k+1)}$.
Since the three groups are randomly chosen,
the state $\rho$ is permutation invariant.
Let us denote elements of $ \bF_2^{n_B'}$ by $x=(x_1, \ldots, x_{n_B'})$, etc.
We define operators 
$\sX^x \equiv X^{x_1} \otimes \cdots \otimes X^{x_{n_B'}}$, 
$\sZ^z \equiv Z^{z_1} \otimes \cdots \otimes Z^{z_{n_B'}}$,
on $(\bC^2)^{\otimes n_B'}$, 
which satisfy
\begin{eqnarray}
\sX_B^x \otimes \sZ_W^{-x}
|G'\rangle
=|G'\rangle, \quad
\sX_W^x \otimes \sZ_B^{-x}
|G'\rangle
=|G'\rangle. \label{H2}
\end{eqnarray}
In the following, 
we regard 
$\sX_B^x$, $\sZ_B^z$ as operators on ${\cal K}_B$
and 
$\sX_W^x$, $\sZ_W^z$ as operators on ${\cal K}_W$.
Here, we distinguish $x$ and $-x$ so that 
we can easily extend our analysis to the qudit case.

Furthermore, for $\bx=(x^1, \ldots, x^{2k+1})
\in (\bF_2^{n_B'})^{2k+1}$
and $\bz=(z^1, \ldots, z^{2k+1})\in (\bF_2^{n_B'})^{2k+1}$, 
using the operator $\sW_B^{x,z}\equiv\sX_B^x \sZ_B^z$ 
on ${\cal K}_B$,
we define
$
\sW_B^{\bx,\bz}\equiv
\sW_B^{x^1,z^1} \otimes \cdots \otimes \sW_B^{x^{2k+1},z^{2k+1}} 
$
on ${\cal K}_B^{\otimes 2k+1}$.
Also, we define $\sW_W^{x,z}$ 
on ${\cal K}_W$,
and $\sW_W^{\bx,\bz}$ on ${\cal K}_W^{\otimes 2k+1}$,
in the same way.
Eq.~(\ref{H2}) implies that
$
\sW_B^{\bx,\bz} \otimes \sW_W^{-\bz,-\bx}
|G'\rangle^{\otimes 2k+1}
=
|G'\rangle^{\otimes 2k+1}
$.
Hence,
\begin{eqnarray*}
&& \Tr \Big[(\sW_B^{\bx,\bz} \otimes \sW_W^{-\bz,-\bx})^{\dagger}
\rho (\sW_B^{\bx,\bz} \otimes \sW_W^{-\bz,-\bx})
|G'\rangle\langle G'|^{\otimes 2k+1}\Big] \\
&= &
\Tr \Big(\rho |G'\rangle\langle G'|^{\otimes 2k+1}\Big).
\end{eqnarray*}
Thus, the discrete-twirled state 
\begin{eqnarray*}
\overline{\rho}\equiv
\sum_{\bx,\bz}
2^{-2 n_B'(2k+1)}
(\sW_B^{\bx,\bz} \otimes \sW_W^{-\bz,-\bx})^{\dagger}
\rho (\sW_B^{\bx,\bz} \otimes \sW_W^{-\bz,-\bx})
\end{eqnarray*}
satisfies 
$
\Tr (\overline{\rho}
|G'\rangle\langle G'|^{\otimes 2k+1} )
=
\Tr (\rho |G'\rangle\langle G'|^{\otimes 2k+1})
$ \cite{BDSW}.
Also, we have
\begin{eqnarray*}
\Tr_1\Big[ (\Tr_{2,3} \overline{\rho})
|G'\rangle\langle G'|^{\otimes k} \Big]
&=&
\Tr_1\Big[ (\Tr_{2,3} \rho) |G'\rangle\langle G'|^{\otimes k}\Big], \\
\Tr_2\Big[ (\Tr_{1,3} \overline{\rho})
|G'\rangle\langle G'|^{\otimes k} \Big]
&=&
\Tr_2\Big[ (\Tr_{1,3} \rho) |G'\rangle\langle G'|^{\otimes k}\Big], \\
\Tr_3 \Big[(\Tr_{1,2} \overline{\rho})
|G'\rangle\langle G'|\Big]
&=&
\Tr_3\Big[ (\Tr_{1,2} \rho ) |G'\rangle\langle G'|\Big].
\end{eqnarray*}
Therefore, we have only to consider the discretely twirled version of $\rho$.
Note that the upper subscript of $x$ and $z$ expresses the choice of group,
and the lower subscript of $x$ and $z$ expresses 
the site of the modified graph.

Finally, let us explain the third point.
Since 
$
( \sW_B^{\bx,\bz} \otimes \sW_W^{-\bz,-\bx})
\overline{\rho}
(\sW_B^{\bx,\bz} \otimes \sW_W^{-\bz,-\bx})^{\dagger} 
=
\overline{\rho}$
and $\rho$ is permutation-invariant, 
the state $\overline{\rho}$ is written 
with a permutation-invariant distribution $P$ on $\bF_2^{2 n_B'(2k+1)}$ as \cite{BDSW}
\begin{eqnarray*}
\overline{\rho}=
\sum_{\bx,\bz} P(\bx,\bz) 
\sW_B^{\bx,\bz}|G'^{\otimes 2k+1}\rangle \langle G'^{\otimes 2k+1}|
(\sW_B^{\bx,\bz})^{\dagger}.
\end{eqnarray*}
Then, we define the function $f$ from $(\bF_2^{n_B'})^{2(2k+1)}$
to $(\{0,1\}^2)^{(2k+1)}$ as
$
f:(\bx,\bz) \mapsto (s_1,t_1), \ldots, (s_{2k+1}, t_{2k+1})$, 
where
$s_i:= 
\left\{
\begin{array}{ll}
0 & \hbox{ if } x^{i}= 0 \\
1 & \hbox{ if } x^{i}\neq  0
\end{array}
\right.$ and 
$t_i:= 
\left\{
\begin{array}{ll}
0 & \hbox{ if } z^{i}= 0 \\
1 & \hbox{ if } z^{i}\neq  0.
\end{array}
\right.$
Here, 
$x^{i}$ and $z^{i}$ are elements of $\bF_2^{n_B'}$.
So, $0$ in the above conditions expresses the zero vector in $\bF_2^{n_B'}$ although 
$s_i$ is an element of $\bF_2$.
\par
\noindent We introduce the distributions $\hat{P}( (s_1,t_1), \ldots, (s_{2k+1}, t_{2k+1}))$
on $(\{0,1\}^2)^{(2k+1)}$ as
$\hat{P} := P\circ f^{-1}$.

Once, Bob's operation is given, 
the values $s_1, \ldots,  s_{2k+1}$, $t_1, \ldots,  t_{2k+1}$
are given as random variables
although half of $s_1, \ldots,  s_{2k}$, $t_1, \ldots,  t_{2k}$
can be observed.
To employ the notations of probability theory, 
we express them using the capital letters as 
$S_1, \ldots,  S_{2k+1}$, $T_1, \ldots,  T_{2k+1}$.
Hence, 
$\hat{P}(S_i=0)$ expresses
the probability that 
the $i$-th measurement outcome of $X$ basis of $B$ system
coincides with the prediction by
the $i$-th measurement outcome of $Z$ basis of $W$ system. 
So, to show Theorem~\ref{L1}, it is enough to show the following 
theorem.
Similarly,
$\hat{P}(T_{k+i}=0)$ expresses
the probability that 
the $k+i$-th measurement outcome of $X$ basis of $W$ system
coincides with the prediction by
the $k+i$-th measurement outcome of $Z$ basis of $B$ system. 
So, to show Theorem~\ref{L1}, it is enough to show the following 
theorem.

\begin{theorem}
\label{L2v2c}
Assume that
$\alpha > \frac{1}{2k+1}$.
When 
the distribution $\hat{P}$ satisfies
\begin{eqnarray*}
&&
\hat{P}(S_{2k+1}=T_{2k+1}=0| S_{j}=T_{k+j}=0~\mbox{for}~ 1\le j\le k)
\\
& \ge & 
1 -\frac{1}{\alpha(2k+1)},
\end{eqnarray*}
the probability
$\hat{P}( S_{j}=T_{k+j}=0~\mbox{for}~ 1\le j\le k ) 
$ is upper bounded by $\alpha$.
\end{theorem}
In this way, we have reduced the problem to the classical hypothesis testing.
The proof of Theorem~\ref{L2v2c}
is given in Appendix.

\acknowledgements
MH is partially supported by the 
JSPS Grant-in-Aid for Scientific Research (A) No. 23246071 and the National
Institute of Information and Communication Technology
(NICT), Japan. The Centre for Quantum Technologies is
funded by the Singapore Ministry of Education and the National
Research Foundation as part of the Research Centres of Excellence programme.
TM is supported by the JSPS Grant-in-Aid for Young Scientists (B) No.26730003 and 
the MEXT JSPS Grant-in-Aid for Scientific Research on Innovative Areas No.15H00850.

\appendix

\section{Relation to previous works}
Here we discuss relations between our result and previous works.
The hypothesis testing of an entangled state 
by local measurements
was initiated by the paper \cite{H1}.
The paper \cite{H1} treats only the maximally entangled state.
The next paper \cite{H2} derived its asymptotic optimal performance
in the i.i.d. setting.
Then, the papers \cite{H3,H4,H5,H6} extended these results 
to the case of a non-maximally entangled state.
However, these papers consider only the locality condition between two parties.
To apply the hypothesis testing to 
the blind quantum computation based on the measurement based quantum 
computation,
we need the following conditions.
\begin{description}
\item[(1)] The test can be applied to the case of a graph state.

\item[(2)] 
Our quantum operations are restricted to single-qubit measurements.
\end{description}

Since the previous studies \cite{H1,H2,H3,H4,H5,H6} do not satisfy 
both conditions,
we cannot use them.
The papers \cite{p1,p2} satisfy these conditions as the verification of graph states, but
their results assume i.i.d. samples,
which is somehow reasonable in laboratory experiments, but cannot
be accepted in quantum cryptography where a malicious adversary can do
anything. 
That is, we need the following additional requirement.
\begin{description}
\item[(3)] 
We cannot assume i.i.d. samples.
\end{description}
Therefore, these results cannot
be directly used for the verification in blind quantum computing.
Our protocol satisfies all of these conditions.

The verification in blind quantum computing was initiated in 
Ref.~\cite{FK}. The idea of their protocol is to use the trap technique:
Alice secretly hides isolated qubits as traps, and if a trap is changed by
Bob, she can detect his malicious behavior.
Several generalizations of Ref.~\cite{FK} have been 
obtained~\cite{topoveri,Elham,Hadjusek},
but all of them essentially use the same idea, namely,
the trap technique.
Our protocol, on the other hand, uses a completely different 
technique for the verification, i.e., the direct graph state testing,
for the first time.

The graph state verification is also used in the context of
the multiprover interactive proof system. For example, Ref.~\cite{Matt}
gave an elegant protocol that uses a device-independent graph state
verification. However, the results in multiprover interactive proof system
can neither be directly used in blind quantum computing, since
the assumption that provers do not communicate with each other is not
natural in the blind quantum computing.

\section{Analysis of local conversion for a bipartite graph state}
We show how the graph state $|G\rangle$ is converted to 
$|G'\rangle \otimes |+\rangle_{B'} \otimes |+\rangle_{W'}$.
For this purpose, we define the notations more formally.
When the true state is $|G\rangle$,
the outcome of measurement with $X$ basis in the party $B$ 
takes values in the subspace over the finite 
filed $\bF_2$.
We denote the subspace by $V_B$, and denote its dimension by $n_B'$.
The orthogonal complement is denoted by $V_B'$.
Then, the space of the measurement outcome with $X$ basis in the party $B$ is written as 
$V_B \oplus V_B'$.
So, we denote the Hilbert space corresponding to $V_B$ and $V_B'$ by 
${\cal K}_B$ and ${\cal K}_B'$, respectively.
Then, we have ${\cal H}_B={\cal K}_B\otimes {\cal K}_B'$.
We denote the eigenstate corresponding to the eigenvalue all $0$ of $X$'s in ${\cal K}_B'$
by $|+\rangle_{B'}$, which is the graph state with isolated sites with no edge.
Similarly, we define 
$V_W$, $V_W'$, ${\cal K}_W$, ${\cal K}_W'$, $n_W'$ and $|+\rangle_{W'}$ and have
${\cal H}_W={\cal K}_W\otimes {\cal K}_W'$
and $n_W'=n_B'$.
Then, we define the graph state $|G'\rangle$ on ${\cal K}_B \otimes {\cal K}_W$ whose graph is composed of isolated edges.
In the following, we show that 
the true state is given as the state $|G'\rangle \otimes |+\rangle_{B'} \otimes |+\rangle_{W'}$.

For an invertible $n_B \times n_B$ matrix $C$, 
we define the unitary operator $U_{C,Z,B}$
as
\begin{align*}
U_{C,Z,B}:=\sum_{z \in \bF_2^{n_B}} |Cz\rangle_B ~_B\langle z|.
\end{align*}
Using the $X$ basis states $|x\rangle_{X,B}$, 
we define the unitary operator $U_{C,X,B}$
as
\begin{align*}
U_{C,X,B}:=
\sum_{x \in \bF_2^{n_B}} |Cx\rangle_{X,B} ~_{X,B}\langle x|.
\end{align*}
Then, we have the relation
\begin{align*}
U_{C,X,B}=U_{(C^{-1})^T,Z,B},
\end{align*}
which can be shown as follows.
\begin{align*}
& U_{C,X,B}| z\rangle_B \\
=&\frac{1}{2^{n_B/2}}
\sum_{x \in \bF_2^{n_B}} |Cx\rangle_{X,B} ~_{X,B}\langle x|
\sum_{x'\in \bF_2^{n_B}} (-1)^{x'\cdot z}
| x\rangle_B
\\
=& \frac{1}{2^{n_B/2}}
\sum_{x \in \bF_2^{n_B}} 
(-1)^{x\cdot z}
|Cx\rangle_{X,B} 
\\
=&
 \frac{1}{2^{n_B/2}}
\sum_{x' \in \bF_2^{n_B}} 
(-1)^{C^{-1} x'\cdot z}
|x'\rangle_{X,B} 
=|(C^{-1})^T z\rangle_B.
\end{align*}
Similarly, we can define 
$U_{D,X,W}$ and $U_{D,X,W}$ for an 
invertible $n_W \times n_W$ matrix $D$.

Next, given the graph state $|G\rangle$, we define the 
$n_B \times n_W $ matrix $A=(a_{i.j})$ as follows.
When the site $j$ of $W$ is connected to the site $i$ of $B$, $a_{i,j}$ is $1$.
Otherwise, $a_{i,j}$ is zero.
Then, we have
\begin{align*}
|G\rangle 
=
\frac{1}{2^{n_W/2}}
\sum_{z \in \bF_2^{n_W}}
|Az\rangle_{X,B} |z\rangle_{W}.
\end{align*}
Then, when we measure $W$ with the $Z$ basis and obtain the outcome $z$,
we obtain $Az$ with the $X$ measurement on $B$.
Then, we have
\begin{align*}
|G\rangle=
\frac{1}{2^{n_B/2}}
\sum_{z \in \bF_2^{n_B}}
|z\rangle_{X,B} |A^{T}z\rangle_{W}
\end{align*}
because any vector $z' \in \bF_2^{n_B}$ satisfies
\begin{align*}
&~_{B}\langle z'|G\rangle \\
=&
\frac{1}{2^{n_B/2}}
\sum_{x \in \bF_2^{n_B}}
~_{X,B}\langle x| (-1)^{x \cdot z'}
\frac{1}{2^{n_W/2}}
\sum_{z \in \bF_2^{n_W}}
|Az\rangle_{X,B} |z\rangle_{W} \\
=&
\frac{1}{2^{n_B/2}}
(-1)^{Az \cdot z'}
\frac{1}{2^{n_W/2}}
\sum_{z \in \bF_2^{n_W}}
 |z\rangle_{W} \\
=&
\frac{1}{2^{n_B/2}}
|A^T z'\rangle_{X,W}.
\end{align*}

Now, we choose a basis 
$c_1, \ldots, c_{n_B'}$
of the $n_B'$-dimensional subspace $V_B$ that is composed of the possible outcomes with $X$ basis in $B$.
Also, we choose $n_B-n_B'$ vectors
$c_{n_B'+1}, \ldots, c_{n_B} $ of $ \bF_2^{n_B}$ such that
$c_1, \ldots, c_{n_B}$ form a basis of $ \bF_2^{n_B}$.
We choose 
$n_W'$ vectors
$d_{1}, \ldots, d_{n_W'} $ of $\bF_2^{n_W}$ such that
$c_i=A d_i $,
and choose a basis 
$d_{n_W'+1}, \ldots, d_{n_W} $ of the kernel of $A$.
Then, we define 
the invertible $n_B \times n_B$ matrix $C$
and
the invertible $n_W \times n_W$ matrix $D$
by
\begin{align*}
C&=(c_1 \ldots c_{n_B}) \\
D&=(d_1 \ldots d_{n_W}).
\end{align*}
So, we define the 
$n_B \times n_W $ matrix $A':
=C^{-1}A D$, which is written as
\begin{align*}
A'=\left(
\begin{array}{cc}
I_{n_B'} & 0 \\
0 & 0 
\end{array}
\right).
\end{align*}
So the state $\frac{1}{2^{n_W/2}} \sum_{z \in \bF_2^{n_W}}
|A'z\rangle_{X,B} |z\rangle_{W} $
corresponds to the graph with many isolated edges and many isolated sites.
Hence, it can be regarded as 
the state $|G'\rangle \otimes |+\rangle_{B'} \otimes |+\rangle_{W'}$.

Now, we have
\begin{align*}
&
\frac{1}{2^{n_W/2}}
\sum_{z \in \bF_2^{n_W}}
|A'z\rangle_{X,B} |z\rangle_{W} \\
=&
\frac{1}{2^{n_W/2}}
\sum_{z \in \bF_2^{n_W}}
|C^{-1}A Dz\rangle_{X,B} |z\rangle_{W} \\
=&
\frac{1}{2^{n_W/2}}
\sum_{z \in \bF_2^{n_W}}
|C^{-1}A z\rangle_{X,B} |D^{-1}z\rangle_{W} \\
=&
U_{C^{-1},X,B} U_{D^{-1},Z,W}
\frac{1}{2^{n_W/2}}
\sum_{z \in \bF_2^{n_W}}
|A z\rangle_{X,B} |z\rangle_{W} \\
=&
U_{C^{-1},X,B} U_{D^{-1},Z,W}
|G\rangle \\
=&
U_{C^{T},Z,B} U_{D^{T},X,W}
|G\rangle \\
=& \frac{1}{2^{n_B/2}}
\sum_{z \in \bF_2^{n_B}}
| z\rangle_{B} | A^T z\rangle_{X,W} \\
=& \frac{1}{2^{n_B/2}}
\sum_{z \in \bF_2^{n_B}}
| C^{T} z\rangle_{B} | D^{T} A^T z\rangle_{X,W} .
\end{align*}
Hence, 
$|G\rangle$
can be converted to 
the state 
$|G'\rangle \otimes |+\rangle_{B'} \otimes |+\rangle_{W'}$
via the local unitary 
$U_{C^{-1},X,B} U_{D^{-1},Z,W}$.
When we apply 
the $Z$ measurement on $W$ and
the $X$ measurement on $B$,
the application of the unitary $U_{C^{-1},X,B} U_{D^{-1},Z,W}$
is equivalent with
the application of the classical conversion 
$x \mapsto C^{-1}x$ and 
$z \mapsto D^{-1}z$.
Similarly, 
when we apply 
the $Z$ measurement on $B$ and
the $X$ measurement on $W$,
the application of the unitary $U_{C^{-1},X,B} U_{D^{-1},Z,W}$
is equivalent with
the application of the classical conversion 
$x \mapsto D^T x$ and 
$z \mapsto C^T z$.
Hence, applying these classical data conversions,
we can treat the state $|G\rangle$
as the state 
$|G'\rangle \otimes |+\rangle_{B'} \otimes |+\rangle_{W'}$.

\section{Concrete examples}
Here, for a better understanding, we demonstrate the above results
for few-qubit graph states.
\subsection{Three-qubit graph state}
Let us consider the case of 
the three-qubit graph state on $\bullet	- \circ - \bullet$.
We number the left black circle $1$ and do the right black circle $2$.
Then, the matrix $A=
\left(
\begin{array}{c}
1 \\
1
\end{array}
\right)$.
Then, $c_1=
\left(
\begin{array}{c}
1 \\
1
\end{array}
\right)$.
Now, we have two choices for $c_2$,
$\left(
\begin{array}{c}
1 \\
0
\end{array}
\right)$
and
$\left(
\begin{array}{c}
0 \\
1
\end{array}
\right)$.
Then, we choose 
$c_2=
\left(
\begin{array}{c}
1 \\
0
\end{array}
\right)$, which implies that
$C=
\left(
\begin{array}{cc}
1 & 1 \\
1 & 0
\end{array}
\right)$
and 
$C^{-1}=
\left(
\begin{array}{cc}
0 & 1 \\
1 & 1
\end{array}
\right)$.
Also, we have $D=1$.
Therefore, the required classical data conversions are given as follows.
When we obtain $X_1$ and $X_2$ as
the $X$ measurement on $B$, and $Z_1$
the $Z$ measurement on $W$,
we need to use the data 
$X_2$, $X_1+X_2$, $Z_1$ instead of the original data.
That is, we check whether the relation $X_2=Z_1$ holds.
When we obtain $Z_1$ and $Z_2$ as
the $Z$ measurement on $B$, and $X_1$
the $X$ measurement on $W$,
we need to use the data 
$Z_1+Z_2$, $Z_1$,  $X_1$ instead of the original data.
That is, we check whether the relation $X_1=Z_1+Z_2$ holds.

\subsection{Four-qubit graph state}
Next, we consider the four-qubit graph state on
the graph $\bullet	- \circ - \bullet- \circ$.
We number the left black circle $1$ and do the right black circle $2$.
Similarly, we number the white circles.
Then, the matrix $A=
\left(
\begin{array}{cc}
1 & 0\\
1 & 1
\end{array}
\right)$.
Now, we choose 
$c_1=
\left(
\begin{array}{c}
1 \\
1
\end{array}
\right)$
and
$c_2=
\left(
\begin{array}{c}
0 \\
1
\end{array}
\right)$, which implies that
$C=A$
and 
$C^{-1}=
\left(
\begin{array}{cc}
1 & 0 \\
1 & 1
\end{array}
\right)$.
Also, we have 
$D=D^{-1}=I$.
Therefore, the required classical data conversions are given as follows.
When we obtain $X_1$ and $X_2$ as
the $X$ measurement on $B$, and $Z_1$ and $Z_2$
the $Z$ measurement on $W$,
we need to use the data 
$X_1+X_2$, $X_2$, $Z_1$, and $Z_2$ instead of the original data.
That is, we check whether 
the relations $X_1+X_2=Z_1$ and $X_2=Z_2$ hold.
When we obtain $Z_1$ and $Z_2$ as
the $Z$ measurement on $B$, and $X_1$ and $X_2$
the $X$ measurement on $W$,
we need to use the data 
$Z_1$, $Z_1+Z_2$,  $X_1$, and $X_2$  instead of the original data.
That is, we check whether the relations $X_1=Z_1$ 
and $X_2=Z_1+Z_2$ hold.

\section{Analysis of the classical hypothesis testing problem}
\label{a2}
In this appendix, we show 

\begin{lemma}\label{LL}
When 
the distribution $\hat{P}$ satisfies
\begin{eqnarray*}
\hat{P}( S_{j}=T_{k+j}=0~\mbox{for}~ 1\le j\le k ) 
\ge &
\alpha,
\end{eqnarray*}
we have
\begin{eqnarray*}
&&\hat{P}(S_{2k+1}=T_{2k+1}=0| S_{j}=T_{k+j}=0~\mbox{for}~ 1\le j\le k) \\
&\ge &
1 -\frac{1}{\alpha(2k+1)}.
\end{eqnarray*}
Here, $\hat{P}$ is a distribution on $2k+1$ trials, in which, each trial 
consists of two bits. 
Also, it is invariant for permutation of $2k+1$ trials.
\end{lemma}
Since Lemma \ref{LL}
is the contraposition of Theorem 2 in the body,
it is sufficient to show Lemma \ref{LL}.

To address permutation-invariant distributions on $2k+1$ trials,
we prepare the typical permutation-invariant distribution 
$\hat{P}_{a,b,c}$, in which, 
the possible numbers of events  
$(0,0)$, $(1,0)$, $(0,1)$, and $(1,1)$
are fixed to 
$2k+1-(a+b+c)$, $a$, $b$, and $c$, respectively.
Hence, the real numbers $a$, $b$ and $c$ satisfy that
$a,b,c \ge 0$ and $2k+1 \ge a+b+c$.
Then, an arbitrary permutation-invariant distribution $\hat{P}$
is written as
\begin{align}
\hat{P}=\sum_{a,b,c} Q(a,b,c) \hat{P}_{a,b,c},\label{25-a}
\end{align}
where $Q$ is a distribution on the set $\{(a,b,c)|a,b,c \ge 0, a+b+c\le 2k+1\}$.
In the following discussion, 
we consider the properties of typical permutation-invariant distributions
$\hat{P}_{a,b,c}$ with the above condition for $(a,b,c)$.

Consider the case when $c \ge 2$. 
Since we detect "1" at least once among $2k$ outcomes
$S_{1}, \ldots ,S_{k},T_{k+1},\ldots,T_{2k}$,
we have
\begin{align*}
\hat{P}_{a,b,c}(S_{j}=T_{k+j}=0~\mbox{for}~ 1\le j\le k )
=0.
\end{align*}

Consider the case when $c =1$. 
When $a \ge k+1$, 
we detect "1" at least once among $k$ outcomes $T_{k+1},\ldots,T_{2k}$ of $X$ basis.
Hence, we have
\begin{align}
& \hat{P}_{a,b,1}(S_{j}=T_{k+j}=0~\mbox{for}~ 1\le j\le k )
=0.\label{24-8}
\end{align}
When $b \ge k+1$, we can show \eqref{24-8} in the same way.
Assume that $a ,b \le k$.
To realize $S_{j}=T_{k+j}=0$ for $1\le j\le k$, 
the following conditions are required.
$a$ events $(1,0)$ occur from $k+1$-th trial to $2k$-th trial.
$b$ events $(0,1)$ occur from $1$-th trial to $k$-th trial.
One event $(1,1)$ occurs in $2k+1$-th trial.
Hence, $k-a$ events $(0,0)$ occur from $k+1$-th trial to $2k$-th trial.
$k-b$ events $(0,0)$ occur from $1$st trial to $k$-th trial.
In this case, the number of total cases is 
$\frac{(2k+1)!}{(2k-a-b)! a! b!}$.
The number of cases satisfying the above conditions is
$\frac{k!}{(k-a)! a! } \frac{k!}{(k-b)! b! }$.
Hence,
we have
\begin{align}
& \hat{P}_{a,b,1}(S_{j}=T_{k+j}=0~\mbox{for}~ 1\le j\le k )\nonumber \\
=&
\frac{\frac{k!}{(k-a)! a! } \frac{k!}{(k-b)! b! }}
{\frac{(2k+1)!}{(2k-a-b)! a! b!}}
=
\frac{k!k! (2k-a-b)!}{(2k+1)!(k-a)!(k-b)!} 
\nonumber \\
=&
\frac{1}{2k+1}\cdot 
\frac{k\cdots (k-a+1)\cdot k\cdots (k-b+1)}{(2k)\cdots (2k+1-a-b)} 
\nonumber \\
\le &
\frac{1}{2k+1} \label{22-5}.
\end{align}
When 
$S_{j}=T_{k+j}=0$ for $1\le j\le k$,
the event $(1,1)$ occurs in $2k+1$-th trial, i.e., $S_{2k+1}=T_{2k+1}=1$.
Thus,
\begin{align*}
& \hat{P}_{a,b,1}(S_{2k+1}=T_{2k+1}=
S_{j}=T_{k+j}=0~\mbox{for}~ 1\le j\le k ) \\
=&\hat{P}_{a,b,1}(S_{2k+1}=T_{2k+1}=0|
S_{j}=T_{k+j}=0~\mbox{for}~ 1\le j\le k ) \\
=&0 .
\end{align*}

Consider the case when $c =0$. 
When $a \ge k+2$ or $b \ge k+2$, 
similar to \eqref{24-8}, we can show that
\begin{align*}
& \hat{P}_{a,b,0}(S_{j}=T_{k+j}=0~\mbox{for}~ 1\le j\le k )
=0.
\end{align*}

Assume that $a ,b \le k+1$.
The condition
$S_{j}=T_{k+j}=0$ for $1\le j\le k$
holds when one of the following three sets of conditions holds.
\begin{description}
\item[(1)]
$b-1$ events $(0,1)$ occur from $1$st trial to $k$-th trial.
$a$ events $(1,0)$ occur from $k+1$-th trial to $2k$-th trial.
The event $(0,1)$ occurs in $2k+1$-th trial.

\item[(2)]
$b$ events $(0,1)$ occur from $1$st trial to $k$-th trial.
$a-1$ events $(1,0)$ occur from $k+1$-th trial to $2k$-th trial.
The event $(1,0)$ occurs in $2k+1$-th trial.

\item[(3)]
$b$ events $(0,1)$ occur from $1$st trial to $k$-th trial.
$a$ events $(1,0)$ occur from $k+1$-th trial to $2k$-th trial.
The event $(0,0)$ occurs in $2k+1$-th trial.
\end{description}
The numbers of cases of (1), (2), and (3) are
${k \choose (b-1)} {k \choose a}$,
${k \choose b} {k \choose (a-1)}$,
and
${k \choose b} {k \choose a}$, respectively.
Since the number of total cases is 
$\frac{(2k+1)!}{(2k+1-a-b)! a! b!}$.
Hence, we have
\begin{widetext}
\begin{align}
& \hat{P}_{a,b,0}(S_{j}=T_{k+j}=0~\mbox{for}~ 1\le j\le k ) 
=
\frac{
{k \choose (b-1)} {k \choose a}
+{k \choose b} {k \choose (a-1)}
+{k \choose b} {k \choose a}
}
{\frac{(2k+1)!}{(2k+1-a-b)! a! b!}}\nonumber
\\
=&
\frac{(2k+1-a-b)! a! b!}{(2k+1)!}
(k!)^2
(
\frac{1}{a!(b-1)!(k-a)!(k-b+1)!}
+\frac{1}{(a-1)! b!(k-a+1)!(k-b)!}
+\frac{1}{a!b!(k-a)!(k-b)!}
) \nonumber\\
=&
\frac{(2k+1-a-b)! a! b!}{(2k+1)!}
(k!)^2
(
\frac{b (k-a+1)}{a!b!(k-a+1)!(k-b+1)!}
+\frac{a (k-b+1)}{a! b!(k-a+1)!(k-b+1)!}
+\frac{(k-a+1)(k-b+1)}{a!b!(k-a+1)!(k-b+1)!}
) \nonumber\\
=&
\frac{(2k+1-a-b)! a! b!}{(2k+1)!}
(k!)^2
\frac{b (k-a+1)+a (k-b+1)+ (k-a+1)(k-b+1)}{a!b!(k-a+1)!(k-b+1)!}
\nonumber\\
=&
\frac{(2k+1-a-b)! a! b!}{(2k+1)!}
(k!)^2
\frac{(k+1)^2- ab}{a!b!(k-a+1)!(k-b+1)!}
\nonumber\\
=&
\frac{(2k+1-a-b)! a! b!}{(2k+1)!}
\frac{((k+1)!)^2}{(k+1)^2}
\frac{(k+1)^2- ab}{a!b!(k-a+1)!(k-b+1)!}
\nonumber\\
= &
\frac{ (k+1)^2-ab}{(k+1)^2}
\frac{(k+1)\cdots (k+2-a)\cdot (k+1) \cdots (k+2-b)}{(2k+1)\cdots (2k+2-a-b)}.
\label{22-6}
\end{align}
\end{widetext}
Here, one might consider that $(k+1)^2$ can be canceled as a common factor.
However, it is not true when $a$ or $b$ is zero.
To keep the validity even in this case, we need to keep the term $(k+1)^2$
in the denominator.

Next, we proceed to $ \hat{P}_{a,b,0}(S_{2k+1}=T_{2k+1}=0,
S_{j}=T_{k+j}=0~\mbox{for}~ 1\le j\le k )$.
The conditions $S_{1}=\ldots =S_{k}=T_{k+1}=\ldots=T_{2k}=0$
holds when the conditions (3) holds.
Hence, we have
\begin{align}
& \hat{P}_{a,b,0}(S_{2k+1}=T_{2k+1}=
S_{j}=T_{k+j}=0~\mbox{for}~ 1\le j\le k )
\nonumber \\
=&
\frac{
{k \choose b} {k \choose a}
}
{\frac{(2k+1)!}{(2k+1-a-b)! a! b!}}
\nonumber\\
=&
\frac{(2k+1-a-b)! a! b!}{(2k+1)!}
\frac{(k!)^2}{a!b!(k-a)!(k-b)!}
 \nonumber\\
=& \frac{k!k! (2k+1-a-b)!}{(2k+1)!(k-a)!(k-b)!}
\nonumber \\
=&
\frac{k \cdots (k+1-a)\cdot k\cdots (k+1-b)}
{(2k+1)\cdots (2k+2-a-b)} \label{22-8}.
\end{align}
Thus,
\begin{align}
& \hat{P}_{a,b,0}(S_{2k+1}=T_{2k+1}=0|
S_{j}=T_{k+j}=0~\mbox{for}~ 1\le j\le k )
\nonumber \\
=&
\frac{
(k+1-a)(k+1-b)
}
{
(k+1)^2-a b
} .
 \label{23-1}
\end{align}

In the following, 
we discuss the distribution $Q$ instead of $\hat{P}$ because of \eqref{25-a}.
Due to the above calculation, 
the event $S_{j}=T_{k+j}=0$ for $1\le j\le k$ occurs
only when $c = 0,1$.
To evaluate the probability of this event, it is sufficient to consider 
the case of $c =0,1$. 
That is, we can restrict the support of the distribution $Q$ to the case of $c =0,1$.
Hence, using a parameter $\beta \in [0,1]$ and 
a distribution $Q_0$ 
on the support $\{(a,b)| a+b \le 2k+1, 0 \le a\le k+1, 0 \le b\le k+1\}$
and a distribution $Q_1$ 
on the support $\{(a,b)| 0 \le a\le k, 0 \le b\le k\}$,
the distribution $Q$ is written as
\begin{align}
Q(a,b,c)=
\left
\{
\begin{array}{ll}
\beta Q_0(a,b) &\hbox{ if }c=0
\nonumber\\
(1-\beta) Q_1(a,b)&\hbox{ if }c=1
\nonumber\\
0 &\hbox{otherwise}.
\nonumber
\end{array}
\right.
\end{align}

Define
\begin{align}
T_1[Q_0]
:=& \sum_{(a,b)} Q_0(a,b)
\hat{P}_{a,b,0}(S_{j}=T_{k+j}=0~\mbox{for}~ 1\le j\le k ) \nonumber\\
T_2[Q_1]
:=&
\sum_{(a,b)}
Q_1(a,b)
\hat{P}_{a,b,1}(S_{j}=T_{k+j}=0~\mbox{for}~ 1\le j\le k ), \nonumber
\end{align}
and
\begin{align}
&T_3[Q_0] \nonumber \\
:=&
\sum_{(a,b)} Q_0(a,b) 
\hat{P}_{a,b,0}
\left(
\begin{array}{l}
S_{2k+1}=T_{2k+1}=S_{j}=T_{k+j}=0 \\
\mbox{for}~ 1\le j\le k 
\end{array}
\right).
\nonumber
\end{align}

Then, we can show the following lemma.
\begin{lemma}\label{L2v2}
Assume that
\begin{align}
\alpha > \frac{1}{2k+1}
 \label{22-2v2b}.
\end{align}
When
\begin{align}
\beta T_1[Q_0]+(1-\beta) T_2[Q_1]
\ge &
\alpha.
 \label{22-2v2}
\end{align}
we have
\begin{align*}
\frac{\beta T_3[Q_0]}{\beta T_1[Q_0]+(1-\beta) T_2[Q_1]}
\ge 
1 -\frac{1}{\alpha(2k+1)}.
\end{align*}
\end{lemma}
Since $\hat{P}(S_{j}=T_{k+j}=0~\mbox{for}~ 1\le j\le k) 
=
\beta T_1[Q_0]+(1-\beta) T_2[Q_1]$
and
$\hat{P}(S_{2k+1}=T_{2k+1}=0| S_{j}=T_{k+j}=0~\mbox{for}~ 1\le j\le k)
=
\frac{\beta T_3[Q_0]}{\beta T_1[Q_0]+(1-\beta) T_2[Q_1]}$,
Lemma \ref{L2v2} yields Lemma \ref{LL}.
Hence, it is sufficient to show Lemma \ref{L2v2}.

The tightness of Lemma \ref{L2v2}, i.e., that of Lemma \ref{LL},
can be shown as follows.
Assume that $\alpha < \frac{1}{2k+1}$,
$\beta=0$, and $Q_1(a,b)=\delta_{a,0}\delta_{b,0}$,
which corresponds to the distribution $\hat{P}$ satisfying the following.
The event $(1,1)$ occurs only in one event, and 
the remaining $2k$ events are $(0,0)$.
Then, Eq.~$\eqref{22-2v2}$ holds
nevertheless 
$\frac{\beta T_3[Q_0]}{\beta T_1[Q_0]+(1-\beta) T_2[Q_1]}
=0$.
That is,
the distribution $\hat{P}$ breaks 
the condition 
$\hat{P}( S_{j}=T_{k+j}=0~\mbox{for}~ 1\le j\le k ) 
\le 
\alpha$
nevertheless 
$\hat{P}(S_{2k+1}=T_{2k+1}=0| S_{j}=T_{k+j}=0~\mbox{for}~ 1\le j\le k)
=0$.
Hence, the constraint $\alpha > \frac{1}{2k+1}$ is crucial.
This situation corresponds to the following Bob's strategy.
Bob generates $2k$ copies of the true state $|G\rangle $ and inserts only one bad state $ \sW_B^{x,z} |G\rangle$, where $x,z$ are non-zero elements.
Then, Bob can success the cheat with probability $\frac{1}{2k+1}$.

To show Lemma \ref{L2v2}, we prepare the following lemma, which will be shown in the next section.
\begin{lemma}\label{L2}
The relation
\begin{align}
& \frac{
(k+1-a)(k+1-b)
}
{
(k+1)^2-a b
} \nonumber \\
\ge &
1- \frac{1}{2k+1}
\frac{(k+1)^2}{ (k+1)^2-a b }  \nonumber \\
& \times 
\frac{(2k+1)\cdots (2k+2-a-b)}
{(k+1) \cdots (k+2-a)\cdot (k+1) \cdots (k+2-b)}
\label{22-14}
\end{align}
holds for $2k+1 \ge a+b$ and $k+1 \ge a,b\ge 0$.
\end{lemma}

\begin{proofof}{Lemma \ref{L2v2}}
Due to \eqref{22-5}, we have 
\begin{align}
T_2[Q_1]\le \frac{1}{2k+1}. \label{22-b}
\end{align}
Hence,
Condition \eqref{L2v2} and \eqref{22-2v2b} imply that
\begin{align}
\beta \ge 
\frac{\alpha -\frac{1}{2k+1}}{T_1[Q_0] -\frac{1}{2k+1}}\label{22-g}
\end{align}
and
\begin{align}
T_1[Q_0] > \frac{1}{2k+1} . \label{22-g-b}
\end{align}

Due to the relations \eqref{22-6} and \eqref{23-1}, 
Lemma \ref{L2} guarantees that
\begin{align*}
&
\hat{P}_{a,b,0}(S_{2k+1}=T_{2k+1}=0| 
S_{j}=T_{k+j}=0~\mbox{for}~ 1\le j\le k )
 \\
\ge & 
1
-\frac{1}{2k+1}\hat{P}_{a,b,0}
(S_{j}=T_{k+j}=0~\mbox{for}~ 1\le j\le k )^{-1}
\end{align*}
when $a$ and $b$ satisfy the conditions $a+b \le 2k+1$, $0 \le a\le k+1$, and $0 \le b\le k+1$.
Hence,
\begin{align*}
&\hat{P}_{a,b,0}(
S_{2k+1}=T_{2k+1}= S_{j}=T_{k+j}=0~\mbox{for}~ 1\le j\le k ) \\
\ge & 
\hat{P}_{a,b,0}(S_{j}=T_{k+j}=0~\mbox{for}~ 1\le j\le k
)-\frac{1}{2k+1}
\end{align*}
under the same condition.
Taking the expectation for $Q_0$, we have
\begin{align}
T_3[Q_0] \ge T_1[Q_0]-\frac{1}{2k+1}.\label{22-a}
\end{align}
Thus, we have
\begin{align*}
& \frac{\beta T_3[Q_0]}{\beta T_1[Q_0]+(1-\beta) T_2[Q_1]} 
\stackrel{(a)}{\ge} 
\frac{\beta T_3[Q_0]}{\beta T_1[Q_0]+\frac{(1-\beta)}{2k+1}} \\
\stackrel{(b)}{\ge} &
\frac{T_3[Q_0]}{\alpha}
\frac{\alpha -\frac{1}{2k+1}}{T_1[Q_0] -\frac{1}{2k+1}} \\
=&
\frac{T_3[Q_0]}{T_1[Q_0] -\frac{1}{2k+1}} 
(1 -\frac{1}{\alpha(2k+1)}) 
\stackrel{(c)}{\ge} 
1 -\frac{1}{\alpha(2k+1)}
\end{align*}
where $(a)$ and $(c)$
follow from \eqref{22-b} and 
the combination of \eqref{22-a} and \eqref{22-g-b}, respectively.
The remaining part $(b)$
follows from \eqref{22-g} and the fact that
$
\frac{\beta T_1[Q_0]}{\beta T_1[Q_0]+\frac{(1-\beta)}{2k+1}} 
=
\frac{ T_1[Q_0]}{T_1[Q_0]+\frac{(\frac{1}{\beta}-1)}{2k+1}} $
is monotone increasing for $\beta$.
\end{proofof}

\section{Proof of Lemma \ref{L2}}
Define
\begin{align*}
& \xi(a,b,k) \\
:=&
{(k+1-a)(k+1-b)}
- (k+1)^2+ab \\
& + \frac{(k+1)^2 \cdot (2k+1)\cdots (2k+2-a-b)}{(2k+1)\cdot
(k+1) \cdots (k+2-a) \cdot (k+1) \cdots (k+2-b) }
\\
=&
\frac{(k+1)^2 \cdot (2k)\cdots (2k+2-a-b)}{
(k+1) \cdots (k+2-a) \cdot (k+1) \cdots (k+2-b) }
\\
&-
(a+b)k +2ab-(a+b).
\end{align*}
Since \eqref{22-14} is equivalent with 
$\xi(a,b,k)\ge 0$,
we show the non-negativity of $\xi(a,b,k)$ in this section.

\subsection{Organization}
Before proceeding to the detailed analysis, we overview the organization of this section.
We firstly show the non-negativity of $\xi(a,b,k)$
for the cases with $k=1,2,3$ in 
Subsections \ref{c1}, \ref{c2}, and \ref{c22}.
In the remaining subsection, we show it 
for the cases with $k\ge 4$.
The detail organization can be summarized as follows.
Without loss of generality, we can assume that $a\ge b$ due to the symmetry of $\xi(a,b,k)$.
Remember that 
the relations
$k+1 \ge a,b \ge 0$ and $2k+1 \ge a+b$
are also assumed.

\subsubsection{Cases: $k=1,2,3$}
Combining discussions in Subsections \ref{c1}, \ref{c2}, and \ref{c22}, 
we can cover all of cases with $k = 1, 2, 3$
due to the following reasons.
The cases with $k=1$
are composed of 
$(a,b)=(0,0),(1,0),(1,1),(2,0),(2,1),(2,2)$.
The cases with $k=2$
are composed of 
$(a,b)=
(0,0),(1,0),(1,1),(2,0),(2,1),(2,2),(3,0),(3,1),(3,2)$.
These cases are covered in Subsection \ref{c1}.

The cases with $k=3$
are composed of 
$(a,b)=
(0,0),(1,0),(1,1),(2,0),(2,1),(2,2),(3,0),(3,1),(3,2)$,
$(3,3),(4,0),(4,1),(4,2),(4,3)$.
In fact, 
the case with $(a,b,k)=(3,3,3)$ is covered in  
Subsection \ref{c22},
and 
the cases with $(a,b,k)=
(4,0,3),(4,1,3),(4,2,3),(4,3,3)$ are covered in  
Subsection \ref{c2}.
So, combining the cases discussed in Subsection \ref{c1},
we can show the cases with $k=3$.
So, we assume that $k \ge 4$ in the following discussion.

\subsubsection{Cases: $a\le 2$ and $k \ge 4$}
The cases with $a \le 2$ are covered in Subsection \ref{c1}.

\subsubsection{Cases: $a+b\le 5$ and $k \ge 4$}
Many cases with $a+b\le 5$ are covered in Subsection \ref{c1}.
The remaining cases are 
$(a,b)=(4,0),(4,1),(5,0)$.
The cases $(a,b)=(4,0),(5,0)$
are covered in in Subsections \ref{c3}, and
the case $(a,b)=(4,1)$
is covered in in Subsections \ref{c4}.

\subsubsection{Cases: $a+b\ge 6$, $a\ge 3$ and $k \ge 4$}
The cases with $a+b\ge 6$, $a\ge 3$ and $k \ge 4$
are classified as Table \ref{t1}.
Hence, all cases have been covered.

\begin{table}[htb]
\caption{Cases with $a+b\ge 6$, $a\ge 3$ and $k \ge 4$}
\label{t1}
\begin{tabular}{|c|c|c|c|}
\hline
Subsection & Condition 1 & Condition 2 & Condition 3\\
\hline
\ref{c5} & \multirow{2}{*}{$a=b$} & $\frac{k+2}{2}\ge a$ & -\\
\cline{1-1}
\cline{3-4}
\ref{c6} &  & $\frac{k+2}{2}< a$ & -\\
\hline
\ref{c7} & \multirow{4}{*}{$a\ge b+1$} & \multirow{3}{*}{$\frac{k+2}{2}\ge b$} & $b\ge 2$
\\
\cline{1-1}
\cline{4-4}
\ref{c3} & & & $b=0 $ \\
\cline{1-1}
\cline{4-4}
\ref{c4} & & & $b=1 $
\\
\cline{1-1}
\cline{3-4}
\ref{c8} & & $\frac{k+2}{2}< b$ & - \\
\hline
\end{tabular}
\end{table}

\subsection{Case: $(a,b)=(0,0),(1,0),(1,1),(2,0),(2,1),(2,2),(3,0),(3,1),(3,2)$}\label{c1}
Since the function $\xi(a,b,k)$ is symmetric for $a$ and $b$, 
we consider only the case when $a\ge b$.
For specific values $(a,b)$, 
$\xi(a,b,k)$ can be calculated as follows.
\begin{eqnarray*}
\xi(0,0,k) &=& \frac{(k+1)^2}{2k+1}, \\
\xi(1,0,k) &=& 0, \\
\xi(1,1,k) &=& 0, \\
\xi(2,0,k) &=& 0, \\
\xi(2,1,k) &=& k-1, \\
\xi(2,2,k) &=& 4 \frac{(k-1)^2}{k} ,\\
\xi(3,0,k) &=& \frac{(k+1)^2}{k-1} ,\\
\xi(3,1,k) &=& 4k-2 ,\\
\xi(3,2,k) &=& \frac{11k^2 -25k +12}{k}.
\end{eqnarray*}
The above values are non-negative when $k \ge a,b$.


\subsection{Case: $a=k+1$}\label{c2}
\begin{align*}
&\xi(k+1,b,k) \\
=&
\frac{(2k)\cdots (k+1-b)}{ k! (k \cdots (k+2-b))}
-(k+1)^2+b(k+1)
 \\
=&
(k+1)
(
\frac{(2k)\cdots (k+1-b)}{ k! ((k+1)\cdots (k+2-b))}
-(k+1-b)
) \\
=&
(k+1) (k+1-b)
( \frac{(2k)\cdots (k+2-b)}{ k! ((k+1)\cdots (k+2-b))}-1
) \\
=&
(k+1) (k+1-b)
( \frac{(2k)\cdots (k+2)}{ k! }-1
) \\
=&
(k+1) (k+1-b)
(\frac{(k+k)(k+k-1)\cdots (k+2)}{ k(k-1)\cdot 2 }
-1
) \\
\ge & 0.
\end{align*}

\subsection{Case: $(a,b)=(3,3)$}\label{c22}
$\xi(3,3,k)$ can be calculated as follows.
\begin{eqnarray}
\xi(3,3,k) &=& 
\frac{2(k-2)}{k(k-1)} (13k^2-29k+12).
\end{eqnarray}
The case $(a,b)=(3,3)$ is possible only when $k\ge 3$.
It is positive in this case.

\subsection{Case: $b=0$ and $a \ge 2$}\label{c3}
Since $\frac{2k \cdots (2k+2-a)}{k\cdots (k+2-a) } \ge 2^{a-1}$,
we have
\begin{align*}
\xi(a,0,k) 
=&
 \frac{2k \cdots (2k+2-a)\cdot (k+1)}{k\cdots (k+2-a) } 
-a (k+1)
\\
=&
(k+1)
(\frac{2k \cdots (2k+2-a)}{k\cdots (k+2-a) } -a ) \\
\ge &
(k+1)
(2^{a-1} -a ) \ge 0 .
\end{align*}

\subsection{Case: $b=1$ and $a\ge 3$}\label{c4}
Since the relations
$\frac{(2k -2)\cdots (2k+1-a)}{(k-1)\cdots (k+2-a) }
\ge 2^{a-2}$ and $\frac{2k-1}{k} \ge 1$ hold, we have
\begin{align*}
&\xi(a,1,k) \\
=&
 2k \frac{2k-1}{k} \frac{(2k-2) \cdots (2k+1-a)}{(k-1)\cdots (k+2-a) } 
-(a+1) k +a-1 \\
\ge &
 2^{a-1}k 
-(a+1) k +a-1 \\
=& (2^{a-1}-a-1)k+a-1 \stackrel{(a)}{\ge} 0,
\end{align*}
where 
$(a)$ follows from 
the inequality $a \ge 3$,
respectively.

\subsection{Case: $a=b$, $a \ge 3$, and $\frac{k+2}{2} \ge a$}\label{c5}
We have 
\begin{align}
& \frac{(2k-1) (2k-2)\cdots (2k+5-2a) (2k+4-2a)}{ ((k-1)\cdots (k+2-a))^2 } \nonumber \\
\ge & 2^{2(a-2)}.\label{24-5}
\end{align}
Since $\frac{k+2}{2} \ge a$, we have
\begin{align}
\frac{(2k+3-2a)}{k}\ge
\frac{(2k+2-2a)}{k}\ge 1
\label{24-6}
\end{align}
Thus, the combination of \eqref{24-5} and \eqref{24-6} yields that
\begin{align}
& \frac{2k \cdots (2k+2-2a)}
{ (k\cdots (k+2-a))^2 } \nonumber \\
=& 
2k \frac{(2k-1) (2k-2)\cdots (2k+5-2a) (2k+4-2a)}{((k-1)\cdots (k+2-a))^2 }\nonumber \\
& \times \frac{ (2k+3-2a) (2k+2-2a)}{ k^2 }
\nonumber \\
\ge &
2^{2(a-2)+1}k.\label{27-a}
\end{align}
Therefore,
\begin{align*}
\xi(a,a,k) 
=& 
\frac{2k \cdots (2k+2-2a)}
{ (k\cdots (k+2-a))^2 } 
-2a k +2 (a^2 -a) \\
\stackrel{(a)}{\ge}  & 
2^{2(a-2)+1}k -2a k +2 (a^2 -a) \\
= &
(2^{2(a-2)+1} -2a)k + 2 (a^2 -a)
\stackrel{(b)}{\ge}  
 0,
\end{align*}
where $(a)$ and $(b)$ follow from 
\eqref{27-a} and the inequality $a \ge 3$,
respectively.

\subsection{Case: $a=b$, $k \ge a> \frac{k+2}{2} $, and $k\ge 4$}\label{c6}
Firstly, we show 
\begin{align}
& \frac{ (2k-1) (2k-2)\cdots 
k}
{ (k(k-1)\cdots (\lfloor \frac{k}{2} \rfloor +1))
\cdot 
( k(k-1)\cdots (\lceil \frac{k}{2} \rceil +1)) } 
\nonumber \\
\ge &  2^{k-2}.\label{24-1}
\end{align}
When $k$ is an even number $2s$,
\begin{align*}
& \frac{ (2k-1) (2k-2)\cdots 
k}
{ (k(k-1)\cdots (\lfloor \frac{k}{2} \rfloor +1))
\cdot 
( k(k-1)\cdots (\lceil \frac{k}{2} \rceil +1)) } 
\\
=&
\frac{ (4s-1) (4s-2)\cdots 
(2s)}
{ ( (s2)(2s-1)\cdots (s +1))^2 } \\
=&
\frac{ (4s-1) (4s-3)\cdots 
(2s +1)}
{ (2s)(2s-1)\cdots (s +1) } 
\frac{ (4s-2) (4s-4)\cdots 
(2s)}
{ (2s) (2s-1) \cdots (s +1) } \\
\ge &
\frac{ (4s-1) (4s-3)\cdots 
(2s +3)}
{ (2s-1)(2s-2)\cdots (s +1) } 
\frac{ (4s-2) (4s-4)\cdots 
(2s+2)}
{ (2s-1)(2s-2) \cdots (s +1) } \\
\ge &
2^{s-1} \cdot 2^{s-1}
= 2^{2s-2}= 2^{k-2}.
\end{align*}
When $k$ is an odd number $2s+1$,
\begin{align*}
& \frac{ (2k-1) (2k-2)\cdots 
k}
{ (k(k-1)\cdots (\lfloor \frac{k}{2} \rfloor +1))
\cdot 
( k(k-1)\cdots (\lceil \frac{k}{2} \rceil +1)) } 
\\
=&
\frac{ (4s+1) (4s)\cdots 
(2s+1)}
{ ( ( (2s+1)(2s)\cdots (s +1) \cdot (2s+1)(2s)\cdots (s +2)) )} \\
=&
\frac{ (4s+1) (4s-1)\cdots 
(2s +1)}
{ (2s+1)(2s)\cdots (s +1)} 
\frac{ (4s) (4s-2)\cdots 
(2s +2)}
{  (2s+1)(2s)\cdots (s +2))} \\
\ge &
\frac{ (4s+1) (4s-1)\cdots 
(2s +3)}
{ (2s)(2s-1)\cdots (s +1)} 
\frac{ (4s) (4s-2)\cdots 
(2s +4)}
{  (2s)(2s-1)\cdots (s +2))} \\
\ge &
2^{s} \cdot 2^{s-1}
= 2^{2s-1}= 2^{k-2}.
\end{align*}
Hence, we obtain \eqref{24-1}.

Considering the even and odd cases in the same way, we can show that
\begin{align}
\frac{(k -1)\cdots(2k+2-2a)}
{ 
(\lfloor \frac{k}{2} \rfloor \cdots (k+2-a))
\cdot 
(\lceil \frac{k}{2} \rceil \cdots (k+2-a)) 
} 
\ge 1.
\label{27-4}
\end{align}
The condition $k \ge a> \frac{k+2}{2}$ implies that 
$a \ge \lceil \frac{k}{2} \rceil +1$.
Hence, we have 
\begin{align}
\lfloor \frac{k}{2} \rfloor +1 \ge k+2-a .
\label{27-3}
\end{align}
Combining \eqref{24-1}, \eqref{27-4}, and \eqref{27-3},
we have
\begin{align}
& 
\frac{2k \cdots (2k+2-2a)}{ (k\cdots (k+2-a))^2 } \nonumber\\
=&
2k \frac{ (2k-1) (2k-2)\cdots 
(k +2)(k +1)k }
{ (k(k-1)\cdots (\lfloor \frac{k}{2} \rfloor +1))
\cdot 
( k(k-1)\cdots (\lceil \frac{k}{2} \rceil +1)) } 
\nonumber \\
& \times 
\frac{(k -1)\cdots(2k+2-2a)}
{ 
(\lfloor \frac{k}{2} \rfloor \cdots (k+2-a))
\cdot 
(\lceil \frac{k}{2} \rceil \cdots (k+2-a)) 
} \nonumber\\
\ge &
2k \frac{ (2k-1) (2k-2)\cdots 
(k +2)(k +1) k }
{ (k(k-1)\cdots (\lfloor \frac{k}{2} \rfloor +1))
\cdot 
( k(k-1)\cdots (\lceil \frac{k}{2} \rceil +1)) } 
\nonumber\\
\ge &
2k 2^{k-2} . \label{24-4b}
\end{align}

Thus,
\begin{align*}
&\xi(a,a,k) 
= \frac{2k \cdots (2k+2-2a)}{ (k\cdots (k+2-a))^2 } 
-2a k +2 (a^2 -a) \\
\stackrel{(a)}{\ge} & 
2k 2^{k-2}
-2a k +2 (a^2 -a)
=
2 (2^{k-2}-a) k
+2 (a^2 -a) \\
\stackrel{(b)}{\ge}  &
2 (2^{k-2}-k) k
+2 (a^2 -a)
\stackrel{(c)}{\ge}  
 0,
\end{align*}
where 
$(a)$, $(b)$, and $(c)$ follow from \eqref{24-4b}, 
the inequality  $k \ge a$, and the inequality $k \ge 4$,
respectively.

\subsection{Case: $\frac{k+1}{2} \ge b \ge 2$ and $a \ge b+1$ and $a+b \ge 6$}\label{c7}
Since the above cases cover all of cases with $a+b \le 5$,
we can assume that $a+b \ge 6$ in the following discussion as well as $k \ge 4$, $b\ge 2 $.
Since $b \ge 2$ and $a \ge b+1 $, we have the following calculation.
\begin{widetext}
\begin{align}
&
2k
\frac{(2k-1) (2k-3)\cdots (2k+3-2b)}{k (k-1)\cdots (k+2-b)}
\frac{(2k-2) (2k-4)\cdots (2k+2-2b)}{k (k-1)\cdots (k+2-b)}
\frac{(2k+1-2b) (2k-2b) \cdots (2k+2-a-b)}{(k+1-b) (k-b)\cdots (k+2-a)}
\nonumber\\
=&
2k
\frac{(2k-1) (2k-3)\cdots (2k+5-2b)}{(k-1)(k-2)\cdots (k+2-b)}
\frac{(2k+3-2b)}{k} 
\frac{(2k-2) (2k-4)\cdots (2k+4-2b)}{(k-1)(k-2)\cdots (k+2-b)} \nonumber\\
&\times \frac{(2k+2-2b)}{k} 
\frac{(2k+1-2b)}{(k+1-b)}
\frac{ (2k-2b) \cdots (2k+2-a-b)}{ (k-b)\cdots (k+2-a)}
\nonumber\\
=&
2k
\frac{(2k-1) (2k-3)\cdots (2k+5-2b)}{(k-1)(k-2)\cdots (k+2-b)}
\frac{(2k-2) (2k-4)\cdots (2k+4-2b)}{(k-1)(k-2)\cdots (k+2-b)}
\frac{(2k+2-2b)}{k} \nonumber\\
&\times \frac{ (2k-2b) \cdots (2k+2-a-b)}{ (k-b)\cdots (k+2-a)}
\frac{(2k+3-2b)}{k} 
\frac{(2k+1-2b)}{(k+1-b)}
\nonumber\\
=&
2k
\frac{(2k-1) (2k-3)\cdots (2k+5-2b)}{(k-1)(k-2)\cdots (k+2-b)}
\frac{(2k-2) (2k-4)\cdots (2k+4-2b)}{(k-1)(k-2)\cdots (k+2-b)}
\frac{(2k+2-2b)}{k} \nonumber\\
&\times \frac{ (2k-2b) \cdots (2k+2-a-b)}{ (k-b)\cdots (k+2-a)}
\frac{(2k+1-2b)}{k} 
\frac{(2k+3-2b)}{(k+1-b)}
\nonumber\\
\stackrel{(a)}{\ge}  &
2 k\cdot 2^{b-2} \cdot 2^{b-2} \cdot 2^{(a-b-1)} \cdot 2
=2^{a+b-3} k, \label{24-4}
\end{align}
\end{widetext}
where $(a)$ follows the following relations.
We have
$\frac{(2k-1) (2k-3)\cdots (2k+5-2b)}{(k-1)(k-2)\cdots (k+2-b)}
\ge 2^{b-2}$,
$\frac{(2k-2) (2k-4)\cdots (2k+4-2b)}{(k-1)(k-2)\cdots (k+2-b)}
= 2^{b-2}$,
$\frac{(2k+2-2b)}{k} \ge 1$,
$\frac{(2k+1-2b)}{k} \ge 1$, and
$\frac{(2k+3-2b)}{(k+1-b)}\ge 2$.
Also, the assumption guarantees the relation $a \ge 2$, which implies that
$\frac{ (2k-2b) \cdots (2k+2-a-b)}{ (k-b)\cdots (k+2-a)}
\ge 2^{(a-b-1)}$.

Hence,
\begin{align*}
\xi(a,b,k) 
=& 
\frac{2k \cdots (2k+2-a-b)}{ (k\cdots (k+2-a))\cdot (k\cdots (k+2-b)) } \\
&-(a+b)k +(2ab -a-b)
\\
=& 
 \frac{2k \cdots (2k+2-a-b)}{ (k\cdots (k+2-a))\cdot (k\cdots (k+2-b)) } \\
&-(a+b)k +(2ab -a-b)
\\
\stackrel{(b)}{\ge} & 
( 2^{a+b-3} -(a+b))k +(2ab -a-b) \stackrel{(c)}{\ge} 0.
\end{align*}
where 
$(b)$ and $(c)$ follow from \eqref{24-4} and
the inequality  $a+b \ge 6$, 
respectively.

\subsection{Case: $b > \frac{k+1}{2}$, $k \ge a \ge b+1$, and $k \ge 4$}\label{c8}
The condition $b > \frac{k}{2}+1 $ implies $b \ge k+1- \lfloor \frac{k}{2}\rfloor$.
Hence, we have $ \lfloor \frac{k}{2}\rfloor +1 \ge  (k+2-b)$.
Thus, we have
\begin{align*}
& \frac{(2k-1) (2k-3)\cdots (2k+3-2b)}{k (k-1)\cdots (k+2-b)} \\
\ge &
\frac{(2k-1) (2k-3)\cdots (2 \lfloor \frac{k}{2}\rfloor +1)
}{k (k-1)\cdots ( \lfloor \frac{k}{2}\rfloor +1)} \\
=&
\frac{(2k-1) (2k-3)\cdots (2 \lfloor \frac{k}{2}\rfloor +2)
}{ (k-1)(k-2)\cdots ( \lfloor \frac{k}{2}\rfloor +1)} 
\frac{(2 \lfloor \frac{k}{2}\rfloor +1)}{k}
\ge 
2^{k-\lfloor \frac{k}{2}\rfloor -1}.
\end{align*}
Since 
$b \ge k + 1 - \lceil \frac{k}{2}\rceil$ implies 
$( \lceil \frac{k}{2}\rceil +1)
\ge (k+2-b)$,
we have
\begin{align*}
& \frac{(2k-2) (2k-4)\cdots (2k+2-2b)}{k (k-1)\cdots (k+2-b)} \\
\ge &
\frac{(2k-2) (2k-4)\cdots (2\lceil \frac{k}{2}\rceil)}
{k (k-1)\cdots ( \lceil \frac{k}{2}\rceil +1)} \\
=&
\frac{(2k-2) (2k-4)\cdots (2\lceil \frac{k}{2}\rceil+2)}
{(k-1)(k-2)\cdots ( \lceil \frac{k}{2}\rceil +1)}
\frac{(2\lceil \frac{k}{2}\rceil)}{k} 
\ge
2^{k-\lceil \frac{k}{2}\rceil -1}.
\end{align*}
Hence,
\begin{align}
&\frac{2k \cdots (2k+2-a-b)}{ (k\cdots (k+2-a))\cdot (k\cdots (k+2-b)) } \nonumber\\
=&
2k
\frac{(2k-1) (2k-3)\cdots (2k+3-2b)}{k (k-1)\cdots (k+2-b)} 
\nonumber\\
&\times 
\frac{(2k-2) (2k-4)\cdots (2k+2-2b)}{k (k-1)\cdots (k+2-b)}
\nonumber\\
&\times 
\frac{(2k+1-2b) (2k-2b) \cdots (2k+2-a-b)}{(k+1-b) (k-b)\cdots (k+2-a)}
\nonumber\\
\ge &
2k
\frac{(2k-1) (2k-3)\cdots (2k+3-2b)}{k (k-1)\cdots (k+2-b)}
\nonumber \\
& \times \frac{(2k-2) (2k-4)\cdots (2k+2-2b)}{k (k-1)\cdots (k+2-b)}
\nonumber\\
\ge &
2k \cdot 2^{k-\lfloor \frac{k}{2}\rfloor -1}
\cdot 2^{k-\lceil \frac{k}{2}\rceil -1}
\nonumber \\
=&
2^{2k-\lfloor \frac{k}{2}\rfloor -\lceil \frac{k}{2}\rceil-1} k
=
2^{k-1} k.\label{24-3}
\end{align}
Thus, we have
\begin{align*}
& \xi(a,b,k) \\
=&
\frac{2k \cdots (2k+2-a-b)}{ (k\cdots (k+2-a))\cdot (k\cdots (k+2-b)) } 
\nonumber \\
& -(a+b)k +(2ab -a-b)
\\
\stackrel{(a)}{\ge}  & 
2^{k-1} k 
-(a+b)k +(2ab -a-b)
\nonumber\\
= &
(2^{k-1}-(a+b)) k +(2ab -a-b) \\
\stackrel{(b)}{\ge} &
(2^{k-1}-2k+1) k +(2ab -a-b)
\stackrel{(c)}{\ge}  0,
\end{align*}
where 
$(a)$, $(b)$, and $(c)$ follow from \eqref{24-3}, 
the inequality  $a+b \le 2k-1$, and the inequality $k \ge 4$,
respectively.



\begin{thebibliography}{00}

\bibitem{MBQC}
R. Raussendorf and H. J. Briegel,
A one-way quantum computer.
Phys. Rev. Lett. {\bf86}, 5188 (2001).

\bibitem{BFK}
A. Broadbent, J. F. Fitzsimons, and E. Kashefi,
Universal blind quantum computation.
Proc. of the 50th Annual IEEE Sympo. on Found. of Comput. 
Sci. 517 (2009).

\bibitem{Barz}
S. Barz, E. Kashefi, A. Broadbent, J. F. Fitzsimons, 
A. Zeilinger, and P. Walther, 
Demonstration of blind quantum computing.
Science {\bf335}, 303 (2012).

\bibitem{unavoidable}
Of course, there are some unavoidable leakages, such as the upper bound of
the Alice's computing size, etc.

\bibitem{Vedran_composability}
V. Dunjko, J. F. Fitzsimons, C. Portmann, and R. Renner, 
Composable security of delegated quantum computation.
Adv. in Crypt. ASIACRYPT 2014, Lecture Notes in Comput. Sci.
{\bf 8874}, 406 (2014).

\bibitem{MABQC}
T. Morimae and K. Fujii, 
Blind quantum computation for Alice who does only measurements.
Phys. Rev. A {\bf87}, 050301(R) (2013).

\bibitem{Vedran_coherent}
V. Dunjko, E. Kashefi, and A. Leverrier, 
Blind quantum computing with weak coherent pulses.
Phys. Rev. Lett. {\bf108}, 200502 (2012).

\bibitem{DI_Joe}
M. Hajdusek, C. A. Perez-Delgado, and J. F. Fitzsimons,
Device-independent verifiable blind quantum computation.
arXiv:1502.02563

\bibitem{DI_Elham}
A. Gheorghiu, E. Kashefi, and P. Wallden,
Robustness and device independence of verifiable blind quantum computing.
arXiv:1502.02571



\bibitem{FK}
J. F. Fitzsimons and E. Kashefi, 
Unconditionally verifiable blind computation.
arXiv:1203.5217.

\bibitem{topoveri}
T. Morimae,
Verification for measurement-only blind quantum computing.
Phys. Rev. A {\bf89}, 060302(R) (2014).

\bibitem{AKLTblind}
T. Morimae, V. Dunjko, and E. Kashefi, 
Ground state blind quantum computation on AKLT state.
Quant. Inf. Comput. {\bf15}, 0200 (2015).

\bibitem{topoblind}
T. Morimae and K. Fujii, 
Blind topological measurement-based quantum computation.
Nat. Comm. {\bf3}, 1036 (2012).

\bibitem{CVblind}
T. Morimae, 
Continuous-variable blind quantum computation.
Phys. Rev. Lett. {\bf109}, 230502 (2012).

\bibitem{Lorenzo}
V. Giovannetti, L. Maccone, T. Morimae, and T. G. Rudolph,
Efficient universal blind computation.
Phys. Rev. Lett. {\bf111}, 230501 (2013).

\bibitem{Joe_intern}
A. Mantri, C. P\'erez-Delgado, and J. F. Fitzsimons, 
Optimal blind quantum computation.
Phys. Rev. Lett. {\bf111}, 230502 (2013).

\bibitem{tri}
Q. Li, W. H. Chan, C. Wu, and Z. Wen, 
Triple-server blind quantum computation using entanglement swapping.
Phys. Rev. A {\bf89}, 040302(R) (2014).

\bibitem{Sueki}
T. Sueki, T. Koshiba, and T. Morimae, 
Ancilla-driven universal blind quantum computation.
Phys. Rev. A {\bf87}, 060301(R) (2013).

\bibitem{distillation}
T. Morimae and K. Fujii, 
Secure entanglement distillation for double-server blind quantum computation.
Phys. Rev. Lett. {\bf111}, 020502 (2013).


\bibitem{Carlos}
C. A. Perez-Delgado and J. F. Fitzsimons,
Overcoming efficiency constraints on blind quantum computation.
arXiv:1411.4777

\bibitem{RHG}
R. Raussendorf, J. Harrington, and K. Goyal,
Topological fault-tolerance in cluster state quantum computation.
New. J. Phys. {\bf9}, 199 (2007).

\bibitem{Gross}
D. Gross and J. Eisert,
Novel schemes for measurement-based quantum computation.
Phys. Rev. Lett. {\bf98}, 220503 (2007).

\bibitem{Miyake}
G. K. Brennen and A. Miyake,
Measurement-based quantum computer in the gapped ground state of a two-body
Hamiltonian.
Phys. Rev. Lett. {\bf101}, 010502 (2008).



\bibitem{Vazirani}
B. W. Reichardt, F. Unger, and U. Vazirani,
Classical command of quantum systems.
Nature. {\bf496}, 456 (2013).



\bibitem{Matt}
M. McKague,
Interactive proofs for BQP via self-tested graph states.
arXiv:1309.5675

\bibitem{BarzNP}
S. Barz, J. F. Fitzsimons, E. Kashefi, and P. Walther, 
Experimental verification of quantum computation.
Nature Phys. {\bf9}, 727 (2013).

\bibitem{respect_Matt}
To our knowledge, the first paper that uses the direct verification
of the graph state in the client-server context is Ref.~\cite{Matt}.

\bibitem{textbook}
E. L. Lehmann and J. P. Romano,
{\it Testing Statistical Hypotheses.}
Springer Texts in Statistics, Springer (2008).

\bibitem{BDSW}
C. H. Bennett, D. P. DiVincenzo, J. A. Smolin, and W. K. Wootters,
Mixed-state entanglement and quantum error correction. 
Phys. Rev. A, {\bf 54}, 3824 (1996).

\bibitem{circuit1}
F. Magniez, D. Mayers, M. Mosca, and H. Ollivier,
Self-testing of quantum circuits.
Automata, Languages and Programming,
Lecture Notes in Comput. Sci. {\bf4051}, 72 (2006).
\bibitem{circuit2}
W. van Dam, F. Magniez, M. Mosca, and M. Santha,
Self-testing of universal and fault-tolerant sets of quantum gates.
Proc. of the 32nd Ann. ACM Symp. on Theor. of Comput. (STOC2000), 688 (2000).

\bibitem{HIKKO}
M. Hayashi, S. Ishizaka, A. Kawachi, G. Kimura, and T. Ogawa, 
{\em Introduction to Quantum Information Science}, Graduate
Texts in Physics, Springer (2014).


\bibitem{H1}
M. Hayashi, K. Matsumoto, and Y. Tsuda, 
A study of LOCC-detection of a maximally entangled state using 
hypothesis testing,
J. Phys. A: Math. Gen. {\bf39}, 14427-14446 (2006).

\bibitem{H2}
M. Hayashi, 
Group theoretical study of LOCC-detection of maximally 
entangled state using hypothesis testing,
New J. Phys. {\bf11}, 043028 (2009).

\bibitem{H3}
M. Owari and M. Hayashi, 
Two-way classical communication remarkably improves local distinguishability, 
New J. Phys. {\bf10}, 013006 (2008).

\bibitem{H4}
M. Owari, and M. Hayashi, 
Asymptotic local hypothesis testing between a pure bipartite state 
and the completely mixed state,
Phys. Rev. A {\bf 90}, 032327 (2014).

\bibitem{H5}
M. Owari, and M. Hayashi, 
Local hypothesis testing between a pure bipartite state and 
the white noise state,
IEEE Transactions on Information Theory {\bf61}(12), pp.6995-7011 (2015).

\bibitem{H6}
M. Hayashi, and M. Owari, 
Tight asymptotic bounds on local hypothesis testing between a 
pure bipartite state and the white noise state,
IEEE International Symposium on Information Theory (ISIT2015), 
Hong Kong, June 14 - June 19, 2015. pp. 691-695 (2015).


\bibitem{p1}
E. Alba, G. Toth, J. J. Garcia-Ripoll, 
Phys. Rev. A {\bf 82}, 062321 (2010).
\bibitem{p2}
J. Joo, E. Alba, J. J. Garcia-Ripoll, T. P. Spiller, 
Phys. Rev. A {\bf88}, 012328 (2013).



\bibitem{Elham}
A. Gheorghiu, E. Kashefi, and P. Wallden,
Robustness and device independence of verifiable blind quantum computing.
arXiv:1502.02571

\bibitem{Hadjusek}
M. Hajdusek, C. A. Perez-Delgado, and J. F. Fitzsimons,
Device-independent verifiable blind quantum computation.
arXiv:1502.02563


\end{thebibliography}
\end{document}